\documentstyle[preprint,prl,aps]{revtex}
\tightenlines

\begin{document}
\draft
\title{Finite size scaling in the 2D XY-model and generalized universality}
\author{G. Palma, T. Meyer and R. Labb\'e}
\address{Departamento de F\'{i}sica, Universidad de Santiago de Chile\\
Casilla 307, Correo 2, Santiago, Chile}
\date{Received \today}
\maketitle

\begin{abstract}
In recent works \cite{BHP} (BHP), a generalized universality has been proposed,
linking phenomena as dissimilar as 2D magnetism  and turbulence. To test these ideas,
we performed a MC study of the $2D XY$-model. We found that the shape of the
probability distribution function for the magnetization $M$ is non Gaussian and
independent of the system size --in the range of the lattice sizes studied-- below the
Kosterlitz-Thoules temperature. However, the shape of these distributions does
depend on the temperature, contrarily to the  BHP's claim. This behavior is
successfully explained by using an extended finite-size scaling analysis and
the existence of bounds for $M$.
\end{abstract}

\pacs{PACS numbers: 05.10.Ln, 05.50.+q, 64.60.C, 75.30.-m}
\narrowtext

The study of critical phenomena is of great interest not only because it allows
the understanding of a large number of very different physical systems, like the super
fluid Helium three, low temperature superconductors, ferromagnetic-paramagnetic systems, 
turbulent fluids and plasmas, polymers, snow flakes and earthquakes, but also due 
to the existence of scale independence of the fluctuations at the critical temperature. 
In fact, although the underlying inter-molecular forces, responsible for the existence
of phase transitions, have a well-defined length scale, the structures they give rise 
do not. This leads, very close to the critical temperature, to the power-law behavior
of physical quantities, which characterizes universality. The main challenge
of the theory of critical phenomena is to explain how dissimilar systems exhibit the same
critical behavior. Renewed interest in this subject has been raised, because in a seminal
paper \cite{BHP} (BHP), it was argued that turbulence
experiments can be explained in terms of a self-similar structure of
fluctuations, just as in a finite critical system like the harmonic finite
$2D XY$-model ($2D HXY$-model). The starting point of this conjecture was the
observation that the probability distribution function (PDF) of the injected
power fluctuations in a confined shear turbulent flow \cite{Labbe} has the same
shape as the PDF of the magnetization in the $2D HXY$-model. It was
also proposed that this analogy should provide a new application of finite
size scaling in critical systems with experimental consequences. 

In this paper, we report the results of a high precision Monte Carlo study of the full
$2D XY$-model. This computation was carried out over the whole physical range
of temperatures. The magnetic susceptibility was computed and the lattice-shifted
critical temperature was obtained for different lattice sizes. Scaling laws for the
magnetization-temperature ratio were tested. Our results agree with the rigorous findings
of Chung \cite{Chung}. We also found that, below the Berezinskii-Kosterlitz-Thouless
temperature $T_{BKT}$, the shape of the PDF of the magnetization is non-Gaussian and
independent of the lattice size, in agreement with previous results \cite{ABH}. However,
we found that the shape of these distributions do depend on the temperature, contrarily to
the generalized universality claimed in (BHP), who stated that the PDF of magnetization
is independent of both, system size and temperature. Our results can be seen as a powerful
extension of finite-size scaling and phenomenological renormalization of the PDF, suggested
originally by Binder in the context of the Ising model \cite{Binder}, with sligth
modifications introduced by the bounds of $M$. This allows in particular to  understand 
the scaling form of the PDF of the order parameter in the $2D XY$-model as well as in the 
turbulent system.

The universality proposed by (BHP) might go beyond the idea of equivalence classes
in Wilson's renormalization group approach \cite{Wilson}, by including into a generalized
universality class systems sharing the properties of finite size, strong correlations
and self-similarity, even if their space-dimensions are different. 

In \cite{ABH}, the two-dimensional probability distribution
for the magnetization is calculated by means of a Monte Carlo simulation in
the context of the $2D HXY$-model. This model is a further simplification
of the Villain model \cite{Villain}, where the vortex variable $n$ is not a
thermodynamical quantity, but it is constrained to the values $n = -1, 0, 1$.
By using diagrammatic techniques, they showed that this asymmetry could be
the result of three-spin interactions and higher order corrections.

Here, we consider the $2D XY$-model, which describes classical planar spins with nearest
neighbor interactions, with a Hamiltonian given by

\begin{equation}
H=-J\sum_{\langle i,j\rangle }\cos (\theta _{i}-\theta _{j})
\label{Hamilton}
\end{equation}
\noindent
where $J$ is the ferromagnetic coupling constant and $\theta _{i}$ is the
angle of orientation of the unitary spin vector $\overrightarrow{s}_{i}$.
The summation $\langle i,j\rangle$ is over nearest neighbors and the
spins are defined on the sites of a square lattice of lattice size $L$, with
periodic boundary conditions. From hereon the ratio $k_{B}/J$
is set equal to unity throughout the paper. This model undergoes a
remarkable binding-unbinding topological phase transition, such that the
free energy and all its derivatives remain continuous \cite{BKT}, and no
long-range order at low temperatures exists, as stated by the Mermin-Wagner
theorem \cite{MWagner}. This model has been extensively studied through both
numerical and analytical methods \cite{XY-m}. 

Our simulation was performed on a
square lattice of lattice sizes L=10, 12, 16, 22, and 32 respectively. We
estimate the MC sweeps needed for thermalization by plotting some
observables like magnetization and energy. Typically $10^{5}$ MC sweeps were
used to reach thermal equilibrium. For thermal averages we used $5\times
10^{5}$ spin configurations $\alpha _{j}$. Because the $2D XY$-model has a
continuous line of critical points below $T_{BKT}$, special care was taken 
to choose statistical independent configurations to evaluate thermal averages 
of physical observables $X$. This was achieved by computing its normalized
autocorrelation function \cite{Binney}

\begin{equation}
C(K)=\frac{<X_{\alpha _{i}}X_{\alpha _{i+K}}>-<X_{\alpha _{i}}><X_{\alpha
_{K}}>}{<X_{\alpha _{i}}^{2}>-<X_{\alpha _{i}}>^{2}}  \label{corr}
\end{equation}
\noindent
where $X_{\alpha _{i}}$ is the value of $X$ in the configuration $\alpha _{i}$
at the $i-th$ step along the MC-path through the configuration space, and the
average $<\dots>$ was taken over this particular path of configurations
separated by $K$ steps from each other. $C(K)=1$ for $K=0$, but for large
enough $K,C(K)$ drops to zero, which means that these configurations become
totally uncorrelated. We choose $K$ so that $C(K)$ was less that the
recommended value $0.05$ \cite{Sokal}. It is well known that as
a critical system approaches the critical temperature, the decorrelation
time $\tau $ diverges with the power law $\tau \sim \xi ^{z}$, where $\xi $
is the (divergent) correlation length of the system and $z$ is known as the
dynamical critical exponent, which is approximately two for local-flip
algorithms like the Metropolis algorithm. This phenomenon is known as
critical slowing down \cite{swang}. This means in practice a serious limitation
to numerical simulations of critical systems close to a critical point.

In fig. 1 we show MC data for the susceptibility for $L^{2}=256$ spins, as a
function of the temperature. The peak occurs at the value $T_{C}(16)=1.15$,
and corresponds to the temperature at which the correlation length equals L,
which is the standard definition of the critical temperature of a finite
system. We compute also the errors (standard deviations), which become
larger as the critical temperature is approached. Another interesting
feature of these errors is that they are larger below $T_{C}$. This can
probably be explained because of the comparative larger correlations lengths
in this region, which corresponds to a continuous line of critical points
with temperature dependent exponents \cite{ID} in the infinite volume limit.

We computed the critical temperature for the lattice sizes L=10, 12, 16, 22,
32. For L=32 we found an effective transition temperature $T_{C}=1.08$, in
agreement with the value obtained by \cite{BH93}, where the linearized RG
equations for the finite size scaling were used. Fig. 2 shows $T_{C}(L)$ as
a function of $(\ln (L))^{-2}$. The values can be described by the finite
size scaling formula \cite{Chung}

\begin{equation}
T_{c}(L) \approx T_{\infty }+\frac{\pi ^{2}}{4c(\ln L)^{2}}  \label{scaltcri}
\end{equation}
\noindent
where $T_{\infty }$ is the extrapolated value of the critical temperature
for infinite volume. Within a few percents of error, we found that the value
of $T_{\infty }$ agrees with the seemingly exact value $0.892$ of the critical
temperature $T_{BKT}$ of the Berezinskii-Kosterlitz-Thouless phase
transition.

The ratio of the mean magnetization to critical temperature is plotted in
fig. 3 as a function of $\ln (L)$. These values are compatible with a
negative straight line, as suggested by \cite{ABH} in the context of the
harmonic XY-model. The values closer to the origin have larger statistical
errors probably because of the finite size effects, which are proportional
to $\ln (L)$ \cite{Chung,Pinn}.

Finally it should be emphasized that we did not make use of scaling
relations to define physical quantities, like the shifted BKT-temperature
$T^{\ast }(L)$, or temperature at which the renormalized spin-wave stiffness
becomes the universal value $2/\pi $ of the infinite system. This is because
the use of the BKT theory beyond its confirmed validity needs at least
justification, (the scaling region is defined by the inequality
$\mid T-T_{BKT})/T_{BKT} \mid <10^{-2}$, where the renormalization group equations
confidently apply \cite{Cardy}). In fact, we were able to obtain accurate
values for thermal averages and test the scaling equations (see eqn. (\ref
{scaltcri}) and Figs. 2 and 3), in spite of the difficulties of numerical
simulations due to the essential finite-size effects present in this model
\cite{Chung} ($\ln (L)$ dependence of physical quantities), and the very
narrow critical region.

The renewed interest in the PDF of fluctuations of magnetization, $M$, is a
consequence of the observation that similarly shaped PDF arise in
completely different systems. For instance, in \cite{Labbe} it was found that
fluctuations of the injected power in a confined turbulent flow show the
same behavior. In Fig. 4 a plot of $\sigma Q(M)$ as a function of
$(M-<M>)/\sigma$ can be seen, for lattice sizes $L=16$ and $L=32$ at the same
temperature $T=0.70$. Here, $Q$ is the PDF of $M$ and corresponds to $P_{L}(M)$ in
the language of Binder (see discussion below eqn. (\ref{PDF-B})).  These curves have 
similar shapes like those found in turbulence experiments, but only within a reduced 
range of temperatures below $T_{BKT}$. These PDFs can be conveniently compared 
with the universal form $\Pi(y)=K(e^{b(y-s)-e^{b(y-s)}})$ proposed in \cite {BHP}, 
by plotting the ratio $\sigma Q(M;T)/\Pi(M)$ vs. $(M - \langle M \rangle)/\sigma$. 
In Fig. 5, four of such plots are displayed for $T=0.40, 0.80, 0.95$ and $1.05$. 
The upper curves are successively multiplied by factors of 10 for clarity. As can 
be seen, when $T$ is increased, these ratios consistently change, showing the 
dependence of $Q(M;T)$ on the temperature. The rightmost part of the lower curves 
is raised, probably due in part to the upper bound $M=1$. On the other hand, for
$T/=0.95$ we can see that the opposite occurs. This temperature is slightly greater
than $T^{*}$, above which the population of spin vortex pairs begins to increase.
This happens because the system volume occupied by these vortices no longer
contributes to the magnetization, which leads to a depleted probability density.
This can be appreciated in the central and rightmost parts of the two upper curves.
The hills are due to the bounds in the magnetization $0<M<1$. 
At $T=1.05$ this effect is greatly enhanced, and the
leftmost part of the curve shows even more clearly the effect of the lower
bound $M=0$. Concerning turbulent flows, we do not expect this type of bounding
effects in the statistics of injected power. In principle, there are no limits to
the fluctuations of such a quantity, and negative values are not excluded, meaning
that the flow is delivering power to the driving system. Although this
type of events are expected to be very unlike, they are not
forbidden.

The temperature dependence of the PDF is not a surprising result. In fact,
the use of the probability distribution of the order parameter to study finite
size scaling and phenomenological renormalization, has been discussed by Binder
in the context of the Ising model \cite {Binder2}. For
the region $\xi \sim L$ he proposed that the probability distribution
function $P_{L}(M)$ does not depend separately on the three variables $\xi,L,M$,
but only on the two scaled combinations, $L/\xi$ and $M\xi ^{\beta/\nu }$:

\begin{equation}
\label{PDF-B}
P_{L}(M)=\xi ^{\beta /\nu }\widetilde{P}(L/\xi ,M\xi ^{\beta /\nu
})=L^{\beta /\nu }P(L/\xi ,ML^{\beta /\nu }).
\end{equation}

He also argued that in the critical region $\xi \gg L$, $P_{L}(M)$ is no
longer Gaussian. In the scaling region, it is a good approximation to take
$P_{L}(M)$ equal to the PDF proposed by Bramwell {\it et al.} (The standard
deviation $\sigma $ plays the role of $L^{-\beta /\nu }$ in the BHP distribution;
this can be seen by using the standard definitions of the critical exponents and
the relation $\sigma =\sqrt{\frac{T}{L^{2}}\chi }$ ). Nevertheless, and away from
the region defined by $\mid T-T_{BKT} \mid /T_{BKT}<10^{-2}$,
there is a temperature dependence in expresion (4) via the correlation length
$\xi$ for finite size.
It turned out that this dependence is rather weak in the range
$0.5<T<T^{*}(L)$ ($T^{*}(L=16) \sim 0.94 $), but out of this range this
dependence becomes stronger due to the presence of vortices and/or bounds.

In conclusion, we found that the probability distribution function for
the magnetization is indeed independent of the system size, but its shape
happens to vary with the temperature of the system, contrarily to the
generalized universality proposed by Bramwell {\it et al.} \cite {BHP}. This
effect comes from the intrinsic temperature dependence on the first scaled
variable $L/\xi$ in the distribution function proposed by Binder for the
order parameter $M$. Also, there is a contribution coming from the constrained
character of the magnetization.

We thank L. Vergara for helpful suggestions. This work was supported in part
through projects FONDECYT 1980608 and 1990169, and DICYT 04-9631PA and 04-9631LM.

\newpage
{\bf Figure captions}

{\bf Figure 1}:Susceptibility for $L=16$ in the range $0 < T \leq 3$. The
peak at $1.15$ corresponds to the shifted critical temperature.

{\bf Figure 2}: The shifted critical temperature for different lattice sizes
is plotted as a function of the system size.

{\bf Figure 3}: Scaling relation for the magnetization-temperature ratio as
a function of system size.

{\bf Figure 4}: Plots of $\sigma Q(M)$ vs. $(M-<M>)/\sigma $ at $T=0.70$ for
lattice sizes $L=16$ (+) and $L=32$ (*).

{\bf Figure 5}: $\sigma Q(M)/\Pi(M)$ ratios for four values of temperature
(see text).

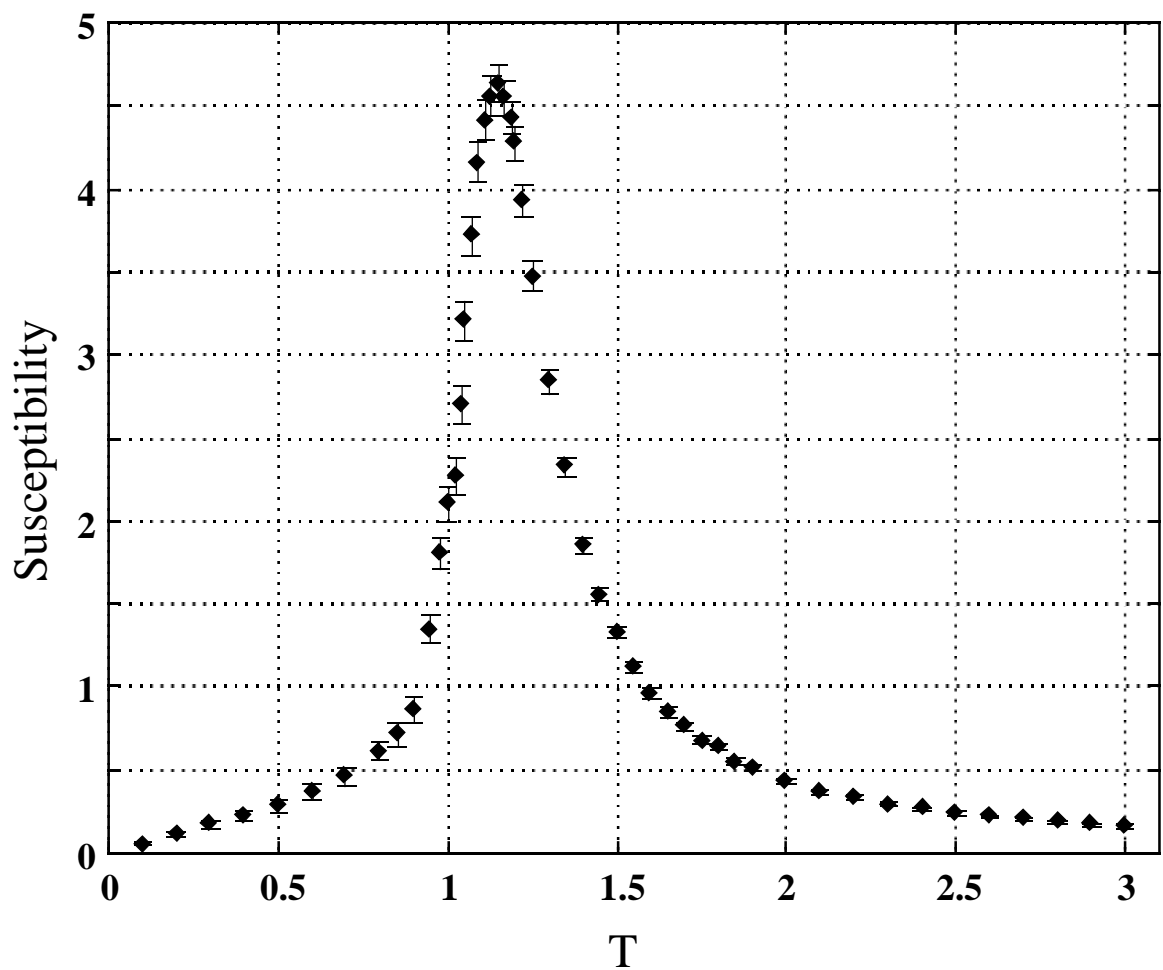
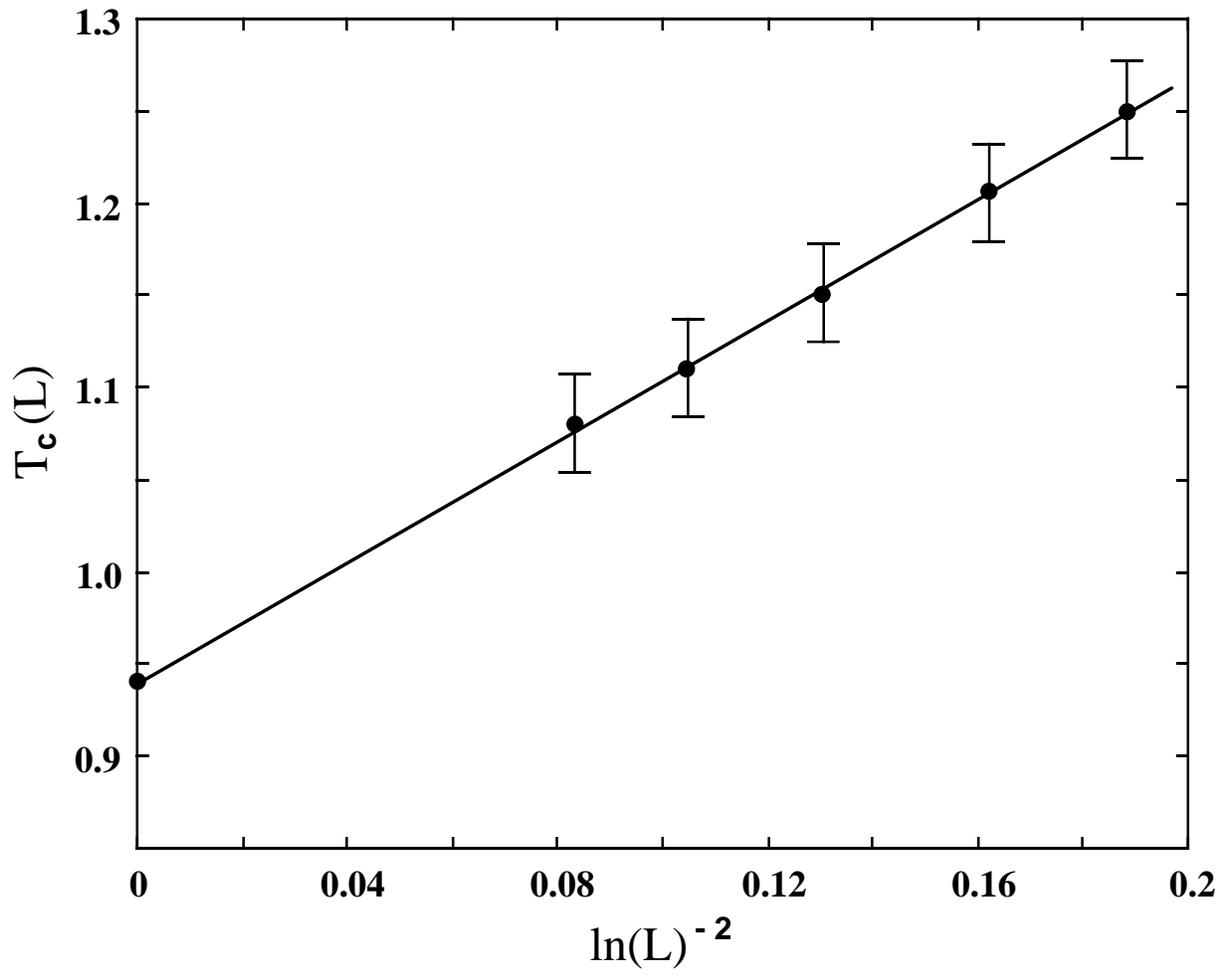
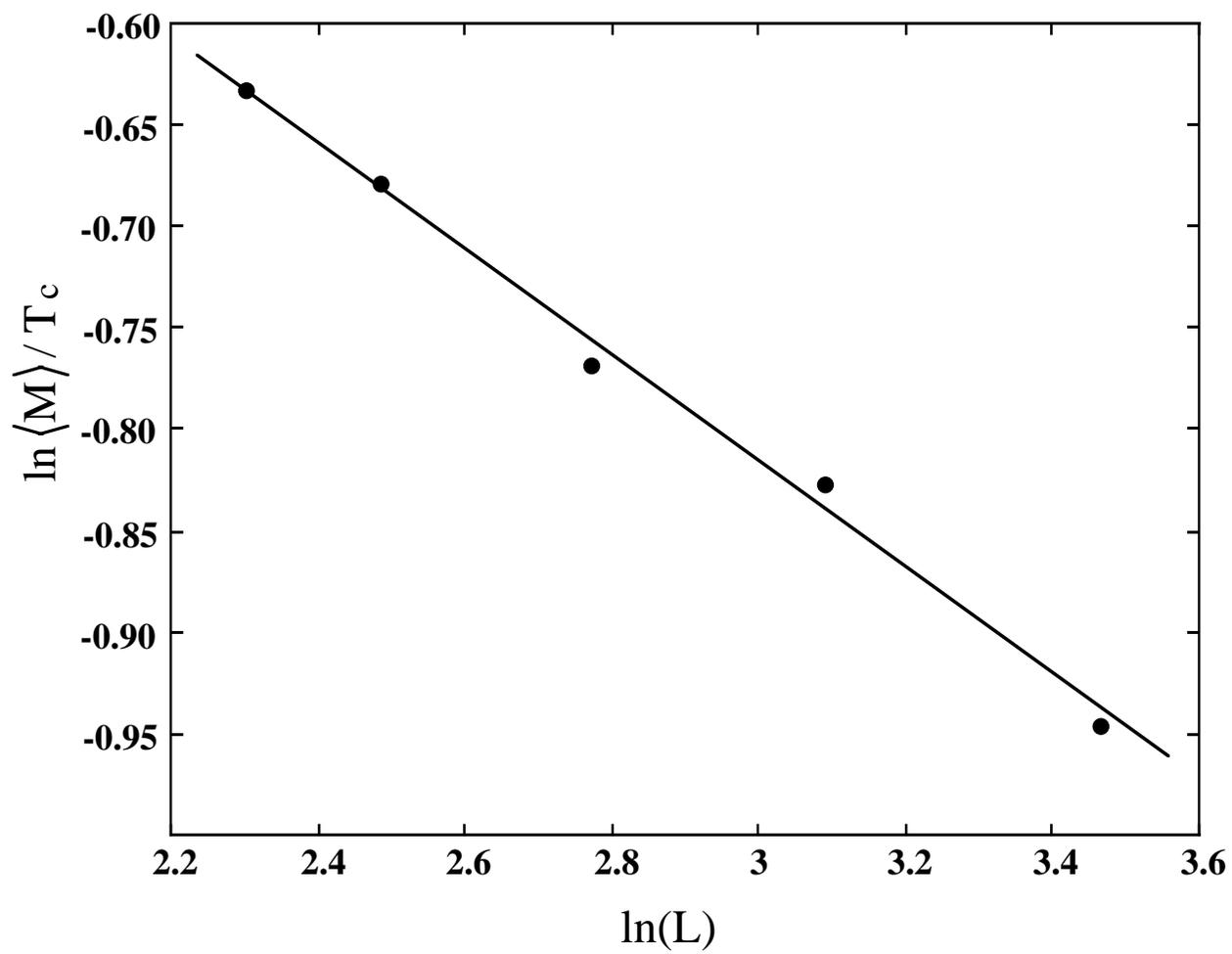
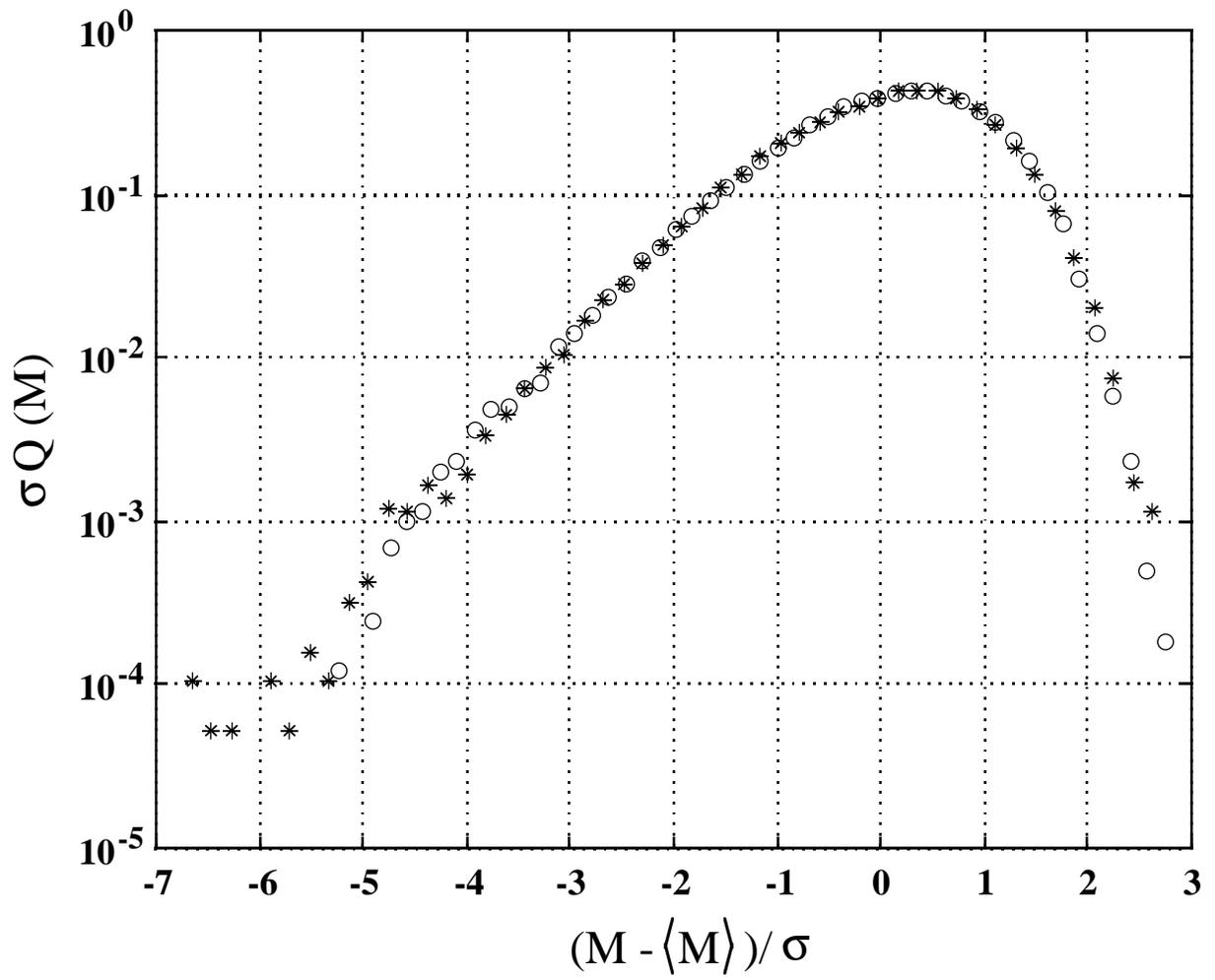
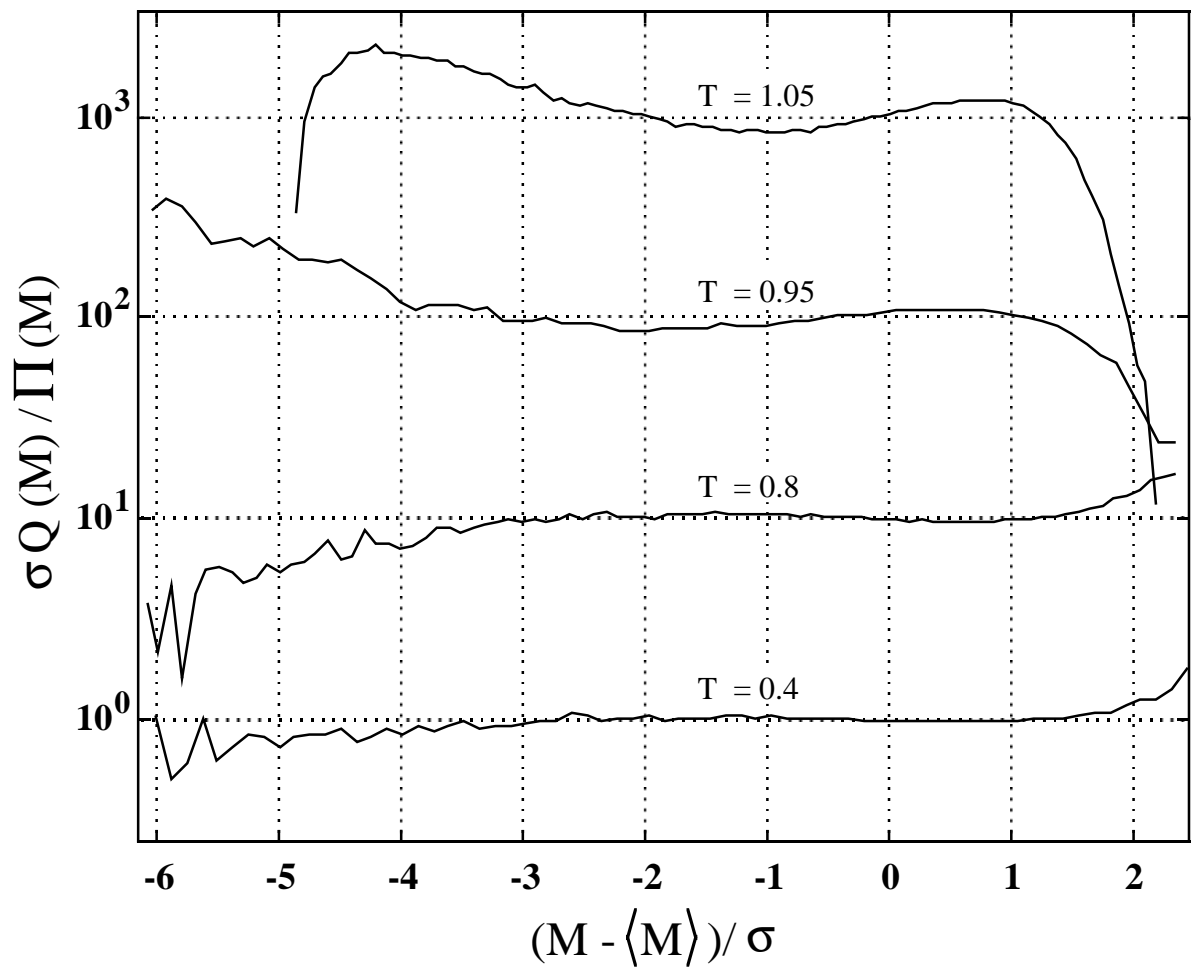
\end{document}